

\magnification=1200 \overfullrule=0pt

\def\i#1{\item{#1}}

\def\b{$$\eqalignno}

\def\to{\rightarrow}
\def\.{\!\cdot\!}
\def\l{\ell}
\def\h{{1\over 2}}
\def\p{\partial}

\def\r2{\sqrt{2}}
\def\bk#1{\langle#1\negthinspace\rangle}
\def\ket#1{|#1\rangle}
\def\bra#1{\langle#1\negthinspace\negthinspace|}

\def\){\right)}
\def\({\left(}
\def\[{\left[}
\def\]{\right]}
\def\.{\cdot}
\input epsf.tex
\centerline{\bf From Multiloop Spinor Helicity Technique
to String Reorganization}\footnote{$^\dagger$}{Talk given
at the Workshop on the Advances in QCD, Minneapolis,

Feb. 18--20, 1994.}
\bigskip
\centerline{C.S. Lam\footnote{*}{email address: Lam@physics.mcgill.ca}}
\medskip
\centerline{Department of Physics,  McGill University}
\centerline{3600 University St., Montreal, P.Q., Canada H3A 2T8}
\bigskip\bigskip
\centerline{\bf ABSTRACT}
\bigskip
The success of the spinor helicity technique for tree processes is reviewed.
To apply it to multiloop diagrams one is naturally led to the Schwinger
proper-time representation, whose properties are discussed. This representation
also serves as a useful link between field theories and string theories.
\bigskip\bigskip \bigskip
\noindent{\bf 1. Introduction}
\bigskip
It is well known that the tree-level cross-section $\sigma_n$ for producing
$n$ {\it soft} photons in the  process
$e^++e^-\to \mu^++\mu^-+\gamma_1+\dots +\gamma_n$ factorizes,
$\sigma_n=\sigma_0\prod_{i=1}^nF(k_i)$, with $\sigma_0$ being the
$e^++e^-\to \mu^++\mu^-$ annihilation cross-section and $k_i$ the momentum
of the $i$th photon. What had not been known until recently$^{1,2}$ was that
such a factorization remained valid for {\it hard} photons all carrying
the same helicity. The {\it spinor helicity technique} was invented$^1$
  to give an easy proof in the case $n=1$. Subsequent refinement$^2$
of the technique made
it possible to prove the factorization for a general $n$, and to apply
it to QCD calculations.
Since then, the technique had been used successfully
to calculate many {\it tree} processes
impossible or too difficult to do by the usual
means.  A good example of this is the Parke-Taylor formula$^3$ so obtained,
which gives a simple and exact expression for the pure gluonic process
$g+g\to g_1+\cdots g_n$ for $n$ produced gluons all carry
the same helicity.
For a review of the subject, see Ref.~4.
\medskip
This technique as discussed could not be  used on higher-order
processes
owing to the presence of loop momenta. This is unfortunate because
loop processes are becoming increasingly important in particle
physics. Happily this difficulty has now been overcome. The first
breakthrough appeared in the one-loop $n$-gluon process, where
superstring techniques$^5$ were used to bypass the loop-momentum problem.
Subsequently it was realized that a first-quantized formalism could lead
to the same result as well$^6$. For arbitrary multiloop processes, these
approaches are ineffective but a purely
field-theoretical technique using the Schwinger proper-time representation
can be devised$^7$.
In the special case of one-loop
pure gluonic processes discussed above, all approaches give identical results.
\medskip
The proper time used in the Schwinger representation is the same
as the worldsheet proper time $\tau$ used in the string theory.
One might therefore expect
Schwinger's proper-time representation to lead to
string-like formulas in field theory, and that is indeed the case.
The formalism of a string theory is very different from that of a
field theory.
In a critical string theory,
every dynamical variable, be it spacetime, color, spin, hypercharge, or
isospin, is treated on an equal footing and each is considered as a function
of the worldsheet variables $\sigma$ and $\tau$. Reparametrization and
conformal invariances of the theory give rise to independence of the scattering
amplitudes on the geometry
of the worldsheet. This in turn leads to duality, `interactions without
interactions', and other magical properties of the string; the low energy
limit of the string theory gives automatically the Maxwell-Einstein-Yang-Mills
gauge theory.  In contrast, ordinary field theory treats the internal
variables like color, spin, and isospin as fields of the spacetime
variables. This asymmetry makes it complicated to enforce
{\it local} gauge invariance, and produces enormous algebraic complications
in actual calculations.
A string-like formalism of field theory can therefore lead to simplifications
and new understandings of gauge theory.
Conversely, field theory is at ease
with multiloop amplitudes with or without external fermions whereas
string theory is not, so the latter may learn a trick or two from the
former also. In short, the Schwinger representation is useful not only
because it allows the spinor helicity technique to be applied to loop
processes, but it also builds a bridge between the vastly different
domains of field theory and string theory, thereby offering an avenue
for cross fertilization.
\medskip
A quick introduction to the spinor helicity technique for tree diagrams
will be given in the next section, and
a brief summary of loop diagrams in the Schwinger representation will be found
in Sec.~3.
\bigskip \bigskip
\noindent{\bf 2. Spinor Helicity Technique for Tree Diagrams}
\bigskip
Spin {\it is} an essential complication in actual calculation:
if relativistic spin-$\h$ particles are present, we must deal with
Dirac matrices and hence
 four-channel problems, for example. This leads to a fair amount
of algebraic complications that fortunately can be sidestepped at
high energies, when the masses of the fermions and other external particles
can be neglected. The method to accomplish that is the {\it spinor
helicity technique}.
\medskip
The physical basis of the simplification is as follows.
For gauge theories, chirality is conserved at the vertices but it is helicity
that is conserved along the propagators. The mixing of the chirality and
helicity eigenstates along the way produces the four-channel problem.
For massless fermions, chirality and helicity are identical,  they
do not change at all from beginning to end,  thus a one-channel problem
results when the fermion masses can be ignored.
\medskip
To be more specific,
let the massless Dirac wave functions be $u_\pm(p)=\pm
v_\mp(p)\equiv\ket{p\pm}$ and
$\bar u_\pm(p)=\pm\bar v_\mp(p)\equiv\bra{p\pm}$. Chirality conservation
implies
$\bk{p_i\pm|p_j\pm}=0$, leaving behind only the  {\it overlap amplitudes}
$\bk{p_i+|p_j-}\equiv [p_ip_j]$ and
$\bk{p_i-|p_j+}\equiv \bk{p_ip_j}$ which do not vanish. In fact,
$$\bk{p_ip_j}[p_jp_i]=2p_i\.p_j,\eqno(1)$$
 and to within a phase factor, both
$[p_ip_j]$ and $\bk{p_ip_j}$ are equal to $\sqrt{2p_i\.p_j}$.
\medskip
Unlike massless fermions, gluon helicities are not conserved, but
even so something can be said about the flow of their helicities.
This is possible  because kinematically a spin-1 particle
can be regarded as a composite of two spin-$\h$ particles.
However, on account of gauge freedom the longitudinal polarization component
is free, so the said composition cannot be unique. Mathematically, this
is reflected in the following representation for the polarization
vector $\epsilon_\mu^\pm$:
$$\epsilon_ \mu ^\pm(p,k)={\bk{p\pm|\gamma _ \mu |k\pm}\over\r2\bk{k\mp|p\pm}},
\eqno(2)$$
where $p$ is the photon/gluon momentum, and $k$ is an {\it arbitrary} massless
momentum called the {\it reference momentum}; different choices of
$k$ correspond to  different choices of its gauge.
\medskip
To see how this helps to visualize the flow of gluon/photon helicity,
and to see how mathematically the four-channel fermion problem can now
be reduced to a one-channel problem, we need two simple mathematical
relations. The first is the completeness relation for massless fermions,
$$\gamma p_i=\ket{p_i+}\bra{p_i+}+\ket{p_i-}\bra{p_i-},\eqno(3)$$
and the second is the Fierz identity which reads
\b{
\bk{A+|\gamma^\mu|B+}\bk{C-|\gamma_\mu|D-}&=2\bk{A+|D-}\bk{C-|B+}&\cr
\bk{A+|\gamma^\mu|B+}\bk{C+|\gamma_\mu|D+}&=2\bk{A+|C-}\bk{D-|B+}.&(4)}$$
We can now understand how the $\gamma$-matrices can be eliminated
in {\it tree diagrams}. Every internal momentum $q_r$ of a tree can
be written uniquely as a linear combination of the external massless momenta
$p_i$,\quad
$$q_r=\sum_i c_{ir}p_i\quad ({\bf trees}).\eqno(5)$$
Using (5) and (3), one can get rid of the $\gamma$-matrices in the
propagators. Using (4) and (2), one can eliminate the $\gamma$-matrices
at the vertices. The disappearance of the $\gamma$-matrices means that the
four-channel problem is now reduced to a one-channel problem. The
resulting scattering
amplitude is a rational function of the overlap amplitudes $[p_ip_j]$
and $\bk{p_ip_j}$.
\medskip
Moreover, one can devise a set of graphical rules to write
down this rational function directly from the Feynman diagram$^7$.
To do so, each photon/gluon line is represented by
a pair of fermion lines, and the resulting fermion lines are all connected
continuously, allowed to terminate only at the external particles of the
diagram. To each fermion line terminating with momenta $p_i$ and $p_j$
is associated an overlap amplitude factor $\bk{p_i\pm|p_j\mp}$.
Fairly obvious rules are available to choose the polarization here.
The numerator of the scattering amplitude is given, up to coupling
constants etc.,
by the product of these factors. The denominators
can also be converted into these factors by using (1).  For example,
the QED diagram Fig.~1(a) should be redrawn as Fig.~1(b) for this purpose.
The $p_i$'s are the particle momenta and the $k_i$'s are the
reference momenta. The signs following the momenta symbols indicate the
helicities. The
numerator of the amplitude can be read off from Fig.~1(b) to be
$$S_0=e^4\.\bk{p_3q_2}[q_2k_6]\.\bk{p_6p_2}
\.[p_4p_5]\.\bk{k_5q_1}[q_1p_1],\eqno(6)$$
where overlap amplitudes involving off-shell momenta are defined
with the help of (5), {\it e.g.,}
$$\bk{p_3q_2}[q_2k_6]\equiv\bk{p_3,p_2-p_6}[p_2-p_6,k_6]
\equiv\bk{p_3p_2}[p_2k_6]
-\bk{p_3p_6}[p_6k_6].\eqno(7)$$
\medskip
If one looks at Fig.~1(b) carefully, one sees that the fermion lines
sometimes turns one way at a vertex, but at other times turns the other
way.  This is all determined by which of the formulas in (4) one is
using. The rule is as follows. Each fermion line has a fixed helicity.
If one moves from one line to the other via a photon, then one continues
along the original direction (opposite direction) if the second fermion
line has the opposite (same) helicity as the first. When it comes to
an external photon line with outgoing momentum $p$, reference momentum $k$, and
helicity $\lambda$, then the rule is as follows. Imagine a fictitious  external
fermion
line with an initial momentum $k$, a final momentum $p$, and helicity $\lambda$
to be attached to the end of this photon line.
Then one can use the rule devised above to proceed between the fermion lines.
\medskip
Similar graphical rules  can be devised for QCD, where one must take into
account color flows in additional to the spin flows discussed here.
\bigskip \bigskip
\noindent{\bf 3. Multiloop Diagrams}
\bigskip
Eq.~(3) is violated for loop diagrams because of the presence of loop
momenta.
To enable the spinor helicity technique to be used one must first
get rid of the loop momenta by integrating them out. This
can indeed be accomplished in the Schwinger-parameter and the Feynman-parameter
representations.
\medskip
In the Schwinger representation,
a proper-time parameter $\alpha$ is introduced for each internal line
of momentum $q$ to convert the denominator of every propagator to
$${1\over -q^2+m^2-i \epsilon}=
i\int_0^\infty d \alpha \exp[-i \alpha (m^2-q^2)].\eqno(8)$$
Suppose the Feynman diagram in question has $\l$ loops with loop
momenta $k_a$, $n$ external lines with outgoing momenta $p_i$,
and $N$ internal lines with momenta $q_r$. Its scattering amplitude
is given by
$$T(p)=\left[ {-i\over
(2\pi)^4}\right]^\l\int\prod_{a=1}^\ell(d^4k_a){S_0(q,p)\over
\prod_{r=1}^N(-q_r^2+m_r^2-i\epsilon)},\eqno(9)$$
where $S_0(q,p)$ consists of the vertices, the numerators of propagators, and
possibly other coefficients. Substituting (8) into (9), the loop integrations
can be carried out, leaving behind the Schwinger-parameter representation$^8$
$$T(p)=\int_0^\infty [D \alpha]\ S(q,p)
\exp\[-iM+iP(\alpha,p)\],\eqno(10)$$
where
\b{
[D\alpha]&={i^N\Delta(\alpha)^{-2}\over (-
16\pi^2)^\l}{\prod_{r=1}^Nd\alpha_r}&\cr
S(q,p)&=  \sum_{k=0}S_k(q,p)&\cr
M&=\sum_r\alpha_rm_r^2&\cr
P&=\sum_r\alpha_rq_r^2.&(11)}$$
There are three points about this new representation to keep in mind.
First, correspondences with the Feynman diagram are maintained, though
loop integrations are now replaced by
$[D\alpha]$, $S_0$ is replaced
by $S$, and the denominators of the propagators $(-q_r^2+m_r^2-i\epsilon)^{-1}$
are replaced by the exponentials $\exp[-i\alpha_r(-q_r^2+m_r^2)]$.
Secondly, the $q_r$ in (10) is defined to be the current flowing through
the $r$th internal line when the Feynman diagram is regarded as an
electric circuit with resistances $\alpha_r$ and external currents $p_i$.
It is no longer the same $q_r$ as in (9);
$k_a$ is no longer present and (5) is restored, though $c_{ir}$
is now dependent on the $\alpha$'s.  Spinor helicity technique can once againn
be applied and a concrete example will be discussed later.
Note that the quantity $P$ in (9) and (10)
is just the power consumed by the circuit.
As such, it is a quadratic form in $p_i$ with coefficients given
by the impedance matrix elements,
$$P=\sum_{i,j}Z_{ij}p_i\.p_j,\eqno(12)$$
though it is important to note that because of momentum conservation,
$\sum_ip_i=0$, $P$ is unchanged under a {\it level transformation}
$$Z_{ij}\to Z_{ij}+\xi_i+\xi_j\eqno(13)$$
so the impedance matrix is not uniquely defined.
In many respects this level transformation resembles a gauge transformation.
Physical quantities such as the power $P$ and the currents $q_r$ are
not altered by this transformation, but
voltage levels at the vertices do change by a {\it common} amount under (13),
which is why the transformation is so named. Measurable quantities
such as voltage differences
are not altered so neither are $P$ nor $q_r$.
As far as (10)--(12) are concerned, any {\it level scheme} (choice of
$\xi_i$ in (13)) of $Z_{ij}$
will give identical results. Later on, we have occasion to see formulas
true only in particular level schemes.
Thirdly, $S=S_0+S_1+S_2+\cdots$ contains the additional terms
$S_k\ (k>0)$, which are obtained from the original $S_0$ by contracting
$k$ pairs of $q$'s via the rule
$$q_r^\mu q_s^\nu\to -{i\over 2}g^{\mu\nu} H_{rs}(\alpha).\eqno(14)$$
If $S_0$ is a polynomial in the $q_r$'s, then $S_k=0$ for $k$
larger than half of its degree, so the sum in $S$ is a finite sum.
\medskip
Simple rules can be derived with the help of graph theory to compute
the electric circuit quantities like current $q_r$ and power $P$, as well as
the {\it Jacobian} $\Delta(\alpha)$ and the {\it contraction functions}
$H_{rs}(\alpha)$:
$$\eqalignno{
\Delta(\alpha )&=\sum_{T_1}(\prod^\ell \alpha ),&\cr
\Delta\. P(\alpha ,p)&=\sum_{T_2}(\prod^{\ell+1}\alpha )
(\sum_1 p)^2,&\cr
-2\Delta\. Z_{ij}&=\sum_{T_2^{ij}}(\prod^{\ell+1}\alpha),\quad ({\rm
zero-diagonal\ level\ scheme}),&\cr
\Delta\. q_r&=\sum_{T_2(r)}\alpha _r^{-1} (\prod^{\ell+1}\alpha
)(\sum_1 p),&\cr
\Delta\. H_{rr}&=-\partial \Delta (\alpha )/\partial \alpha _r,
&\cr
\Delta\. H_{rs}&=\pm\sum_{T_2(rs)}(\alpha _r \alpha _s)^{-1}
(\prod^{\ell+1}\alpha ),\quad(r\not= s).&(15)}$$
These formulas should be interpreted as follows. An $\l$-loop diagram
can be changed into a tree by cutting $\l$ lines, and into two disconnected
tress
(a {\it `2-tree'}) by cutting $\l+1$ lines. The sums in (15) are taken over the
collection $T_1$
of all such trees in the case of $\Delta$, over the collection $T_2$ of all
such 2-trees in the case of $P$, and over the collection
$T_2(r)$ of all 2-trees in which the line $r$ must be cut to form them,
in the case of $q_r$. For $H_{rs}$, the sum is over the
collection $T_2(rs)$
of 2-trees in which lines $r$ and $s$ must be cut, and such that a single tree
results if either of them is inserted back.
For $Z_{ij}$, where the formula is true only in the {\it zero-diagonal level
scheme} in which $Z_{ii}=0$ for all $i$, the sum is taken over the set
$T_2^{ij}$ of 2-trees in which vertices $i$ and $j$ belong to separate
trees.
In all cases, $\prod \alpha$ indicates the product of the $\alpha$'s of the
cut lines, and $\sum_1 p$ denotes the sum of external momenta attached
to either one of the two trees. The signs involved in the formulas
for $q_r$ and $H_{rs}$ can also be determined easily.
\medskip
For example, (15) leads to the following results for the circuit in Fig.~2.
$$\eqalignno{
\Delta&=(\alpha_1+\alpha_2)(\alpha_3+\alpha_4)+\alpha_5(\alpha_1+
\alpha_2+\alpha_3+\alpha_4)&\cr
-2\Delta\. Z_{13}&= (\alpha_1 \alpha_2 \alpha_3 + \alpha_1 \alpha_2
\alpha_4 + \alpha_1 \alpha_3 \alpha_4 + \alpha_2 \alpha_3 \alpha_4) +
\alpha_5(\alpha_1+\alpha_4)(\alpha_2+\alpha_3) &\cr
\Delta\.q_3&=\alpha _5 \alpha _1(p_1+p_2)
+(\alpha _5 \alpha _2)p_2-\alpha _4(\alpha _1+\alpha _2+\alpha _5)p_3&\cr
\Delta\.P&=\alpha _1 \alpha _5 \alpha
_3(p_1+p_2)^2+\alpha_2\alpha_5\alpha_4(p_1+p_4)^2+
\alpha _1 \alpha _2(\alpha _3+\alpha _4+\alpha
_5)p_1^2+&\cr
&+\alpha_3\alpha_4(\alpha_1+\alpha_2+\alpha_5)p_3^2
+\alpha_1\alpha_5\alpha_4p_4^2+\alpha_2\alpha_5\alpha_3p_2^2
&\cr}$$
\medskip
The spin flow for the numerator $S_0$ in (10) can be read off directly from the
Feynman diagram as before. For example, Fig.~3(a) should first be drawn like
Fig.~3(b), from which one gets immediately that
$$S_0=e^4\.\bk{p_4q_2}[q_2p_3]\.[p_2q_1]\bk{q_1q_4}[q_4q_3]\bk{q_3p_1}.$$
Similar rules can be devised for other $S_k$.  In the case of QCD,
there are also analogous graphical rules for color flows.
\medskip
The integrand of the scattering amplitude in (9) is a function of $\alpha$
and $p$, but it is not just any function of them. The integrand is
composed of electric circuit quantities. As such, they obey the
 Kirchhoff laws and a set of differential identities
\b{
{\p P\over\p \alpha_s}&={\p \over\p \alpha_s}\(\sum_r\alpha_rq_r^2\)=q_s^2,&\cr
{\p q_r\over\p \alpha_s}&=H_{rs}q_s,&\cr
{\p H_{rs}\over\p \alpha_t}&=H_{rt}H_{ts}.&(16)}$$
Many other identities can be derived from these.
\medskip
These identities can be used to reshape (9) into other expressions.
In particular, into
a string-like form.
We shall do that only for scalar electrodynamics but similar formulas
are known for QED and QCD. Consider first a one-loop $n$-photon amplitude,
given by Fig.~4(a). In scalar electrodynamics, photons are derivatively
coupled to the charged particles, hence
$$S_0(q,p)=\prod_a\[e\epsilon_a\.(q_{a'}+q_{a''})\]\equiv S_0^{ext}(q,p).
\eqno(17)$$
Using (16), one can transform the integrand of (9) into a form
obtained in string theory$^5$ and the first-quantized formalism$^6$:
$$S(q,p)\exp[-i(M-P)]=\exp[-i(M-P')]_{ml},\eqno(18)$$
where
$$P'=\sum_{a,b}
(p_a-ie\epsilon_a\p_a)\.(p_b-ie\epsilon_b\p_b)Z_{ab},\eqno(19)$$
with $\p_a\equiv\p/\p\alpha_{a'}-\p/\p\alpha_{a''}$,
if
$P=\sum_{a,b}p_a\.p_bZ_{ab}$.
Eq.~(18) is true only in the zero-diagonal level scheme where $Z_{aa}=0$
for every vertex $a$. The subscript $ml$ instructs us to expand the
exponential and keep only the terms  multilinear in all
the $\epsilon_a$'s.
For arbitrary multiloop processes like Fig.~4(b), a similar string-like formula
exists$^7$. In this case
$S_0(q,p)=S_0^{ext}(q,p)S_0^{int}(q,p)$, where $S_0^{ext}$ is the product of
vertex factors at vertices with an external photon line, as in (17),
and $S_0^{int}$ is the rest of the vertices, indicated
by heavy dots in Fig.~4(b). Eq.~(18) now takes on the form
$$S(q,p)\exp[-i(M-P)]=S^{int}(q',p)\exp[-i(M-P')]_{ml},\eqno(20)$$
for some suitably defined $q'$, the detail of which is discussed in Ref.~7.
\bigskip
\bigskip
\centerline{\bf References}
\medskip
\i{1.} F.A. Berends, R. Kleiss, P. De Causmaeker, R. Gastmans, and T.T. Wu,
{\it Phys.
Lett.}  103B (1981), 124; P. De Causmaecker, R. Gastmans, W. Troost and
T.T. Wu, {\it Phys. Lett.} 105B (1981), 215.
\i{2.} Z. Xu, D.H. Zhang, and L. Chang, {\it Tsinghua Univesity preprints}
84/4,5,6 (1984); {\it Nucl. Phys.}  B291 (1987), 392.
\i{3.} S. Parke and T. Taylor, {\it Phys. Rev. Lett.} 56 (1986),
2459.
\i{4.} M.L. Mangano and S.J. Parke, {\it Phys. Rep.}  200 (1991), 301;
R. Gastmans and T.T. Wu, `The Ubiquitous Photon',
International Series of Monographs on
Physics, Vol.~80 (Clarendon Press, Oxford, 1990).
\i{5.} Z. Bern and D.K. Kosower, {\it Phys. Rev. Lett.}  B86 (1991), 1669.
\i{6.} M. Strassler, {\it Nucl. Phys.}  B385 (1992), 145.
\i{7.} C.S. Lam, {\it Nucl. Phys.}  B397 (1993), 143; {\it Phys. Rev.}
{\bf D48} (1993), 873;  {\it Can. J. Phys.} (1994) (hep-ph/9308289).
\i{8.} C.S. Lam and J.P. Lebrun, {\it Nuovo Cimento} 59A (1969), 397.
 \end